\documentclass[pre]{revtex4}
 
\usepackage{graphicx}
\usepackage{epsf}
\usepackage{psfig}
\usepackage{dcolumn}
\usepackage{bm}

\begin{document}

\title{First Passage Time Statistics For Systems Driven by Long Range
Gaussian Noises}

\author{Aldo H. Romero}
\affiliation{Facultad de F\'{\i}sica, Pontificia Universidad Cat\'olica
de Chile, Casilla 306, Santiago 22, Chile}
\author{J. M. Sancho}
\affiliation{Departament d'Estructura i Constituents de la Mat\'eria,
Universitat de Barcelona, Av. Diagonal 647, 08028 Barcelona. Spain}
\author{Katja Lindenberg}
\affiliation{Department of Chemistry and Biochemistry and Institute
for Nonlinear Science, University of California San Diego,
9500 Gilman Drive, La Jolla, California 92093-0340, USA}



\begin{abstract}
We examine the mean first passage time for a particle driven by highly
correlated Gaussian fluctuations to reach one or more predetermined
boundaries.  We discuss a numerical algorithm to generate power-law
correlated fluctuations and apply these  to three physical examples.  One is
the arrival of a free particle at either end of an interval.  The second is
the decay of a particle from an unstable state.  The third is the
time for a particle to cross a barrier separating one well from
another in a double well potential.  In each case a comparison with
the first passage time for a particle driven by Gaussian white noise is
presented, as is an analysis of the dependence of the first passage
time properties on the correlated noise parameters.  
\end{abstract}

\maketitle

\section{Introduction}

The arrival of a random process at a particular state often triggers some
important behavior.  Among the large number of instances that one can name
are the firing of a neuron, the nucleation of a phase associated with a phase
transition, the triggering of an alarm, the occurrence of a major
earthquake, and the crossing of an activation barrier by a reaction
coordinate that converts reactants to products in a chemical reaction.  These
examples range from the macroscopic to the microscopic, indicating that
the arrival problem is of interest on all scales.  In many instances it
is not only important to know the
statistics of occurrence of such events, but more specifically the statistics
of {\em first} occurrence.  This then leads to the study of the statistical
properties of the time that it takes a random process to reach a
specified state for the first time, that is, the {\em mean first passage
time}.  The first passage time literature is enormous and extends over many
decades~\cite{stratonovich,wax,gardiner,risken,redner}.

For simplicity we restrict our attention in this work to one-dimensional
random processes $x(t)$, although the generalization to vector
processes or even to fields is conceptually straightforward.
The statistical properties of the temporal evolution of the random
process are typically described in one of two ways.  One is by way
of an evolution equation for the probability distribution $P(x,t)$ for the
random variable to take on the value $x(t)=x$ at time $t$ (the random
variable and the values that it can take on are often denoted by the same
symbol, $x$ in this case, although the distinction should be kept in
mind). The other is by
way of an evolution equation (``Langevin equation") for the random variable
itself.  From the solution of the appropriate evolution equation 
(via steps that are in general non-trivial) one can then obtain the first
passage time statistics for $x(t)$ to reach a prescribed value for the
first time.  In either case, one often thinks of the evolution of
$x(t)$ as being driven by fluctuations $\eta(t)$ of prescribed
statistical properties from which the statistics of $x(t)$ then follow.  

Let us then turn to the ``driving fluctuations" $\eta(t)$ and their typical
statistical properties.  By far the most common assumption is that
fluctuations have a Gaussian distribution, although there has
always been a great deal of interest in fluctuations that are
distributed in some other fashion.  The most common examples include
L\'{e}vy distributions~\cite{ditlevsen,grigolini1}
(distributions with very long
tails),
and dichotomous processes~\cite{ourbook,grigolini2}
(where the random variable can take on only two values).  The former
typically arise when the fluctuations are a result of many multiplicative
inputs; the latter often serve as an analytically tractable prototype.

In order to specify the fluctuations fully, one must also explicitly state
their correlation properties.  If they are Gaussian,
\begin{equation}
P(\eta)=\frac{1}{\sqrt{2\pi}\sigma} e^{-\eta^2/2\sigma^2},
\label{gaussian}
\end{equation}
then only the average
$\langle \eta(t)\rangle$ (usually and here as well taken to be zero)
and two-time correlation function (assumed stationary)
$\gamma(t-t')\equiv \langle\eta(t)\eta(t')\rangle$ need to be specified,
since all other
correlation properties then follow.  In Eq.~(\ref{gaussian})
$\sigma^2=\gamma(0)$.  For other distributions it is in
general necessary to also specify higher order correlations functions,
even though these are often difficult to determine.  The early
literature dealt primarily with $\delta$- correlated or ``white noise"
processes,
\begin{equation}
\gamma(t)= 2D\delta(t)
\label{delta}
\end{equation}
and $D$ (the integral of the correlation function)
measures the intensity of the noise.  In this case the second
moment $\sigma^2$ diverges as $D/\Delta t$, where
$\Delta t\rightarrow 0$ is
the ``width" of the $\delta$-function and therefore the shortest time scale
in the problem.  The evolution equation $P(x,t)$ for a random process
driven by Gaussian $\delta$-correlated fluctuations $\eta(t)$ is the
familiar Fokker-Planck equation~\cite{stratonovich,risken}, and the
associated
first passage time properties are well
understood~\cite{stratonovich,risken,FPfpt}.

About two decades ago a great deal of attention began to be directed toward
understanding the effects of ``colored noise", that is, of driving
fluctuations that are not $\delta$-correlated.  
Most of the attention was focused on
{\em exponentially correlated} noise,
\begin{equation}
\gamma(t) = \frac{D}{\tau} e^{-|t|/\tau},
\label{exp}
\end{equation} 
where $\tau$ is the correlation time of the noise.  The associated
evolution equation for $P(x,t)$ and related first passage time properties
as a function of the correlation time were studied in detail and are also
well understood~\cite{reviews}.   For instance, it is firmly established
that the mean first passage time from one potential well to another of a
process $x(t)$ driven by exponentially correlated noise increases with
increasing correlation time.  It is also understood that the
particular exponential form of the correlation function is not crucial in
the qualitative features of the first passage time statistics.  The most
important feature determining the qualitative features of the first passage
time statistics is the fact that there is a {\em finite} correlation time
associated with the fluctuations.

More recently there has been considerable interest in
(typically Gaussian) fluctuations that display
long-range power-law correlations with an infinite correlation time.  Such
highly correlated fluctuations have been considered in a broad array of
circumstances ranging from the biological to the physical, from the
economical to the atmospheric, and encompassing theoretical and experimental
studies~\cite{makse96,Eva1}.  There are two separate issues that need to be
addressed with such highly correlated noise.  One is the numerical
generation of the correlated noise itself. Only recently have the
traditional Fourier filtering methods been superseded by far more efficient
procedures~\cite{makse96,nises,pang,Barabasi,Romero99} that will 
briefly be reviewed below in the context of our systems.
The second issue is that of the effect of such a driving noise on the first
passage time properties of the system that is driven by these highly
correlated fluctuations, and this is our principal interest in this
paper.  Herein we examine the mean first passage time of
a particle, driven by highly correlated fluctuations, in three different
physical situations. In the
first case, we discuss the arrival of a free particle 
(moving superdiffusively) at
an absorbing boundary.  As a second case, we study the decay of a particle
from an initial unstable state. Finally, we discuss the
time for a particle
to cross a barrier separating one well from the other in a double well
potential.

In Sec.~\ref{definenoise} we summarize the main steps in the generation of 
long ranged correlated Gaussian noises. In Secs.~\ref{superdiffusion},
\ref{decay}, and \ref{barrier} we present respectively the arrival of the
superdiffusive particle at an absorbing boundary, the decay of
an unstable state, and the barrier crossing
problem. Finally, we conclude with a summary in Sec.~\ref{conclusions}.

\section{Generation of long range correlated Gaussian noise}
\label{definenoise}

The main steps to generate noises with a Gaussian distribution and
arbitrary correlation properties can be found in
Refs.~\cite{makse96,nises,pang,Barabasi,Romero99}. Here
we present a summary of these ``spectral methods" including some details
relevant to our applications.
The goal is to generate a Gaussian noise  $\eta(t)$, 
with correlation function $\gamma (t)$ defined by:
\begin{equation}
\langle \eta (t) \eta (t') \rangle  =  \gamma (t - t'),
\label{corret}
\end{equation}
and with a Fourier transform
\begin{equation}
\widetilde{\gamma}(\omega)  =  \int_{-\infty}^\infty dt\,e^{-i \omega t } 
\gamma(t).
\label{correw}
\end{equation}
This correlation function may be specified analytically or numerically.
In the $\omega$ Fourier-space, the transformed noise
$\widetilde{\eta}(\omega)$
has a  correlation function
\begin{equation}
\langle \widetilde{\eta}(\omega) \widetilde{\eta}(\omega') \rangle  =
2\pi \gamma(\omega) 
\delta( \omega + \omega').
\end{equation}

The algorithm for the noise generation can be summarized 
as follows.  First, the time interval $(0,t)$ is discretized into
$N=2^n$ intervals of mesh 
size $\Delta t$.  This time interval has to be much smaller 
than any other characteristic time of the system (and hence $N$ must be
sufficiently large) because it is used as the time integration step
in the numerical simulation of the stochastic evolution equations.
The intervals in time will be denoted by a roman index and the resulting
frequency intervals in Fourier space are denoted by a greek index. 
\begin{figure}[htb]
\begin{center}
\includegraphics[width=9cm]{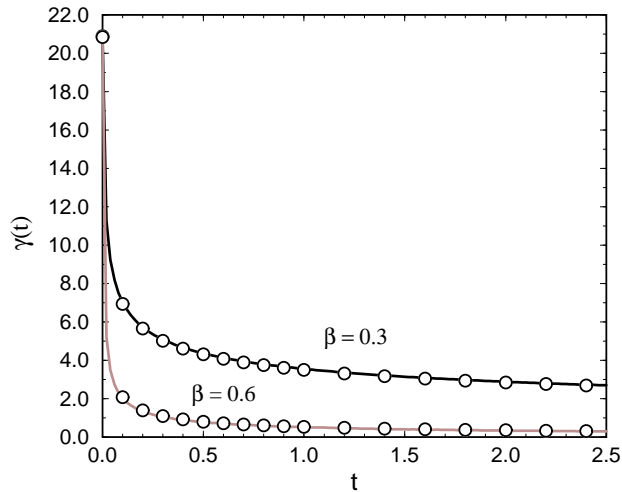}
\end{center}
\caption{Comparison of Eq.~(\protect\ref{powercorr}\protect) (circles) with
correlation functions generated with the algorithm described in
the text (solid lines).
Upper curve: $\beta = 0.3$ with $t_0 = 0.0026$. Lower
curve: $\beta = 0.6$ with $t_0 = 0.0022 $.  
Other parameters for both curves: $\epsilon = 20.0$, 
$N=2 ^{19}, \Delta t = 0.01$, $\omega_0 = 0.0001$. These simulation results
are averaged over 100 realizations. 
}
\label{fig:corrlong1}
\end{figure}
\begin{figure}[htb]
\begin{center}
\includegraphics[width=9cm]{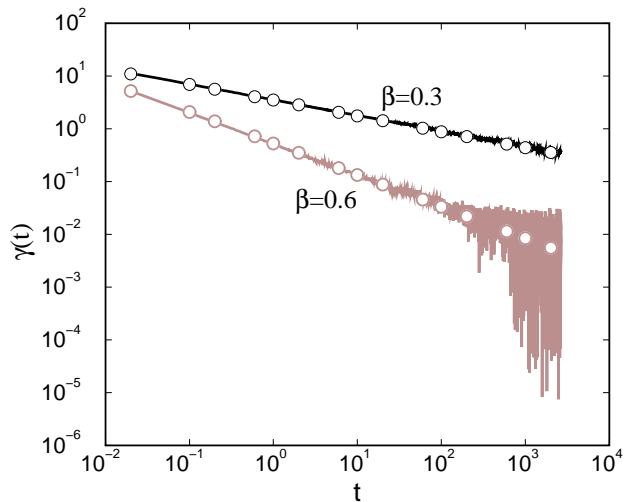}
\end{center}
\caption{Same as Fig.~\protect\ref{fig:corrlong1}\protect but on 
a logarithmic scale that
highlights the power law behavior.
}
\label{fig:corrlong2}
\vspace*{0.1in}
\end{figure}
In discrete Fourier space the noise has a correlation function given by
\begin{equation}
\langle \widetilde{\eta}(\omega_\mu) \widetilde{\eta}(\omega_{\mu'}) \rangle
 =  N \Delta t\widetilde{\gamma}( \omega_{\mu})\delta_{\mu + \mu' , 0},
\end{equation}
where $\widetilde{\eta}(\omega_\mu)$ can be constructed as follows:
\begin{equation}
\begin{array}{l l l}
\widetilde{\eta}(\omega_\mu) =  \sqrt{ N \Delta t
\widetilde{\gamma}(\omega_{\mu)})} \ 
\alpha_{\mu}, & &
\mu  =  0,\ldots,N-1
\nonumber
\\ \\
\widetilde{\eta}(\omega_N)  =  \widetilde{\eta}(\omega_0), & &
\omega_\mu  =  \frac{2 \pi \mu}{N \Delta t}.
\end{array}
\label{noisew}
\end{equation}
Here the $\alpha_{\mu}$ are Gaussian random numbers 
with zero mean and with correlation
\begin{equation}
\langle \alpha_\mu \alpha_\nu \rangle \ = \ \delta_{ \mu +\nu,0}~.
\label{anticor}
\end{equation}
This type of $\delta$-anticorrelated noise can be generated rather easily
if the symmetry properties of real periodic series in the Fourier space 
($\alpha_i$) are used \cite{nises,Romero99}.  Owing to the periodicity
of $\widetilde{\gamma}(\omega_{\mu})$, the index $\mu$ can be made to
run from $-N/2$ to $N/2$, and the maximum frequency is
$\omega_{max} = \pi/\Delta t$ \cite{Romero99}.  The Fourier components are
related by $\alpha_{\mu+pN}=\alpha_\mu$ for any integer $p$, and
$\alpha_{-\mu}=\alpha_\mu^\ast$ because the original random numbers in time
are real.  The anticorrelated random numbers can
then be constructed as $\alpha_\mu = a_\mu +i b_\mu$ with $b_0=0$,
$a_0^2=1$, and
$a_\mu$ and $b_\mu$ are each Gaussian random numbers with zero mean and a
variance of $1/2$ for $\mu\neq 0$.    

The discrete inverse transform of any sequence
$\widetilde{\eta}(\omega_\mu)$ is then numerically 
calculated by a standard Fast Fourier Transform algorithm. The result 
is a string of $N$ numbers, $\eta(t_i)$ which, by construction, have the 
proposed time correlation~(\ref{corret}).  However, due to the symmetries 
of the Fourier transform only $N/2$ of these values are actually 
independent and the remaining numbers are periodically correlated to
them. Thus we have constructed a true  random process from $t=0$ to a
maximum time  $T_{cutoff}= \Delta t N / 2$.

This algorithm is sufficiently general to allow generation of noise of
any given correlation function.  For instance, in~\cite{Romero99} we
considered Gaussian noise with a Gaussian correlation function,
\begin{equation}
\gamma(t)=\frac{2D}{\tau\sqrt{2\pi}} e^{-t^2/2\tau^2}
\label{corr1}
\end{equation}
whose Fourier transform can be obtained analytically,
\begin{equation}
\widetilde{\gamma}(\omega)=2D e^{-\tau^2\omega^2/4}.
\end{equation}
According to our prescription, if we generate a discrete field of random
numbers according to
\begin{equation}
\widetilde{\eta}(\omega_\mu)=
\left[2D N\Delta t \exp\left( \frac{\tau^2}{(\Delta
t)^2}(\cos(2\pi\mu/N)-1)\right)\right]^{1/2} \alpha_\mu 
\label{gen1}
\end{equation}
for sufficiently large $N$ 
the resulting correlation function upon Fourier inversion
should match Eq.~(\ref{corr1}) (it does).
Similarly, for exponentially correlated Gaussian noise, cf.~Eq.(\ref{exp}),
whose Fourier transform is
\begin{equation}
\widetilde{\gamma}(\omega)=\frac{2D}{1+\tau^2\omega^2},
\end{equation}
we generate the discrete random numbers according to
\begin{equation}
\widetilde{\eta}(\omega_\mu)=\left( \frac{2D N\Delta
t}{1+\left(\frac{2\tau}{\Delta t} \sin(\pi\mu/N)\right)^2 }\right)^{1/2}
\alpha_\mu
\label{gen2}
\end{equation}
and again reproduce the correlation function (\ref{exp})
upon Fourier inversion.
In both cases $2D$ is the intensity of the noise and $\tau$ its
correlation time.
We find that it does not matter whether we discretize the $\omega$ using
the function $\cos(2\pi\mu/N)$ as in (\ref{gen1}) or
$\sin(\pi\mu/N)$ as in (\ref{gen2}).

A more difficult noise to generate numerically 
is one with a memory of inverse power law form,
\begin{equation}
\gamma(t) =  \frac {\varepsilon}{ (1 + |t|/t_0)^\beta}. 
\label{powercorr}
\end{equation}
Here $t_0$ is an adjustable (small) parameter. 
For reasons described below and related to our particular algorithm,
the value of $t_0$ turns out to depend on $\beta$.
When the power law decay exponent $\beta >2$ then not only is the
intensity of
the fluctuations finite (and given by $2\varepsilon t_0/(\beta-1)$, which
increases with decreasing $\beta$), but
they also have a finite correlation time (which also grows as $\beta$
decreases.  The effects of this sort of noise are essentially the same as
those of any Gaussian noise with a finite correlation time, although direct
comparison with most of the literature would have to be done
carefully~\cite{reviews} because correlation time effects have usually
been studied in the context of noise of a fixed intensity.  When $1 <
\beta < 2$ then the fluctuations have a finite intensity but an infinite
correlation time.   We have not analyzed this case.
Even more persistent correlations, and the ones of interest to us here,
occur when $0 < \beta < 1$. 

In order to implement the spectral method, we would first have to know the
Fourier transform of the correlation function.  The form (\ref{powercorr})
does not have an closed-form Fourier transform~\cite{note},
and so instead we postulate or guess the form
\begin{equation}
\widetilde{\gamma}(\omega_\mu) = \frac {\varepsilon \beta \pi
\omega_{max}^{-\beta}}
{\left[ \frac {2}
{\Delta t} \sin( \pi \mu / N ) + \omega_0 \right]^{(1 - \beta)}} \, 
\label{eta} 
\end{equation}
where $\omega_{max}$ was defined earlier as $\pi/\Delta t$ and
where $\omega_0$ is a low frequency cutoff that is chosen to control
the low-frequency behavior and avoids a zero-frequency divergence. 
To check whether Eq.~(\ref{eta}) is indeed an appropriate choice one
has to perform the numerical inverse transform and compare with
(\ref{powercorr}).  For a given exponent $\beta$ this comparison
involves the three
parameters $\Delta t$ (the discretization time step), 
$\omega_0$, and $t_0$.  The value of $\omega_0$ that leads to the best
agreement turns out to depend on the time step.  In our calculations we
have mostly used $\Delta t = 0.01$ but in our barrier-crossing calculations
in Sec.~\ref{barrier} we take $\Delta t = 0.02$.  For this range of $\Delta t$
we find that the best choice (giving the best
concordance between the desired correlation function and the assumed
Fourier transform) is $\omega_0 = 0.0001$.  This is the value used
throughout this analysis.

\vskip 30pt

\begin{figure}[htb]
\begin{center}
\mbox{\scalebox{.35}{\includegraphics[angle=-90]{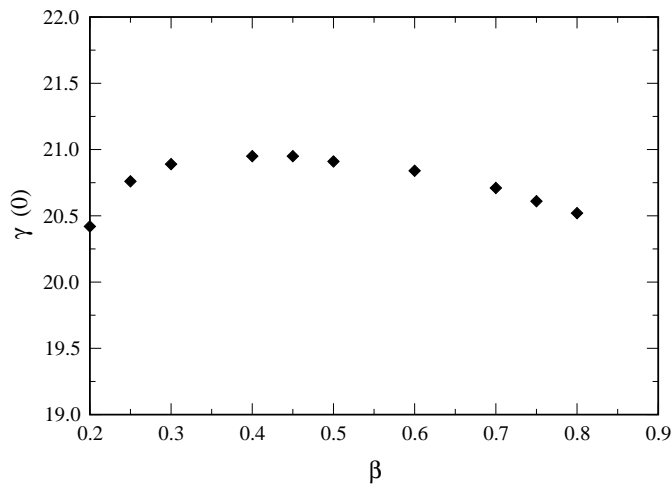}}}
\end{center}
\caption{ $\gamma(0)$ vs $\beta$ obtained from the simulations. 
}
\label{fig:gamma}
\end{figure}            
\begin{figure}[htb]
\begin{center}
\mbox{\scalebox{.35}{\includegraphics[angle=-90]{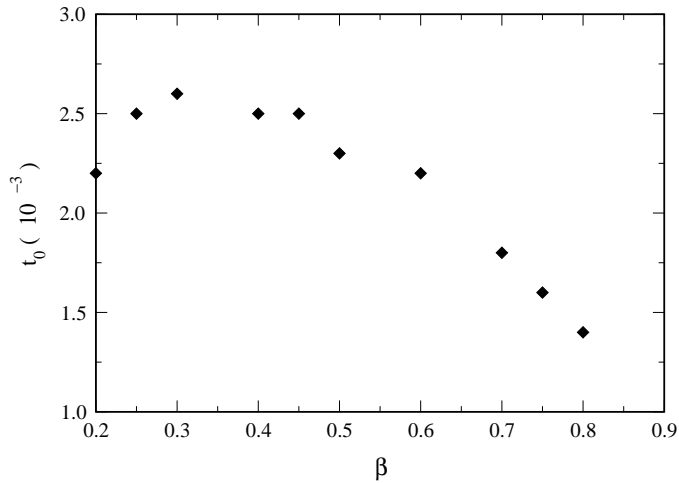}}}
\end{center}
\caption{Parameter $t_0$ found from the best fitting of the
autocorrelation function
Eq.~(\protect\ref{powercorr}\protect).}
\label{fig:t0}
\vspace*{0.1in}
\end{figure}            

From Eq.(~\ref{eta}) we obtain an important relation, 
\begin{equation}
\gamma(0) \simeq \frac{1}{\pi} \int_{0}^{\omega_{max}}
\widetilde{\gamma}(\omega) d \omega = \epsilon + {\cal
O}(\omega_0/\omega_{max}),
\label{gamma0}
\end{equation}
consistent with Eq.~(\ref{powercorr}).
Therefore $\gamma(0)$ is independent of $\beta$ and is a direct measure of
the noise control parameter $\varepsilon$.  This provides a first
check for our algorithm.

A second test of our algorithm is presented in  Figs.~\ref{fig:corrlong1}
and \ref{fig:corrlong2}, where
we exhibit the numerically generated correlation functions as well as the
analytic counterparts with $\beta =0.3$ and $0.6$. Within the resolution
of these figures, the simulated
results agree well with the expected power laws given by Eq.(\ref{powercorr}) 
for the particular parameter choices shown in the caption.  The value of
$\varepsilon$ has been assumed to be selectable independently of $\beta$
and according to Eq.~(\ref{gamma0}).  
Figure~\ref{fig:gamma} shows that this is indeed the case, i.e., that
$\gamma(0)$ as generated from our simulations is essentially independent of
$\beta$.  The small variations are due to the discreteness of the time
variable and the stochasticity of the problem.  More complicated is the
choice of the parameter $t_0$: for a fixed $\varepsilon$, the choice of
$t_0$ that leads to a best match between the numerically generated
correlation function and the analytic form (\ref{powercorr}) depends on
$\beta$, a dependence that is exhibited in 
Fig.~\ref{fig:t0}. The parameter $t_0$ also varies with the time
step $\Delta t$, a dependence we have not exhibited because we hold $\Delta
t$ fixed in our analysis.  In any case, implementation of our algorithm
requires attention to these dependences.  The values of $t_0$ indicated in
Figs.~\ref{fig:corrlong1} and \ref{fig:corrlong1} result from this best fit.

\begin{table}
\begin{center}
\begin{tabular}{|c|c|c|}
\hline
$\beta$   	& Theory & Simulation \\ 
& $\frac{\displaystyle 2}{\displaystyle 2- \displaystyle \beta}$&\\ 
\hline
0.25	 	&   1.143       &     1.151 \\
0.5 	 	&   1.333       &     1.344 \\
0.75	 	&   1.600       &     1.596 \\
\hline
\end{tabular}
\label{table1}
\end{center}
\caption{Comparison of theoretical and simulation exponents}
\end{table}

\begin{figure}[htb]
\vspace*{0.2in}
\centerline{
\mbox{\scalebox{.35}{\includegraphics[angle=-90]{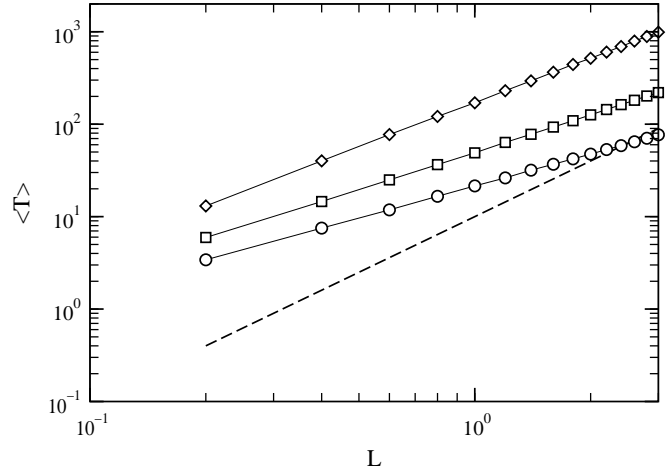}}}}
\caption{
Mean first passage time, $\langle T \rangle$, 
as function of the interval length, $L$, for a free particle
driven by power law correlated
noises with different exponents $\beta$. In all cases
$\varepsilon =  0.05$. The dashed line is the theoretical
prediction for white noise. The lines joining the simulation
symbols are simply guides for the eye.  Circles: $\beta=0.25$; Squares:
$\beta=0.50$; Diamonds: $\beta=0.75$.  The exponents  
in Table~I are fits to these numerical
results.
}
\label{fig:xt}
\vspace*{0.2in}
\end{figure}

\begin{figure}[htb]
\centerline{
\mbox{\scalebox{0.35}{\includegraphics[angle=-90]{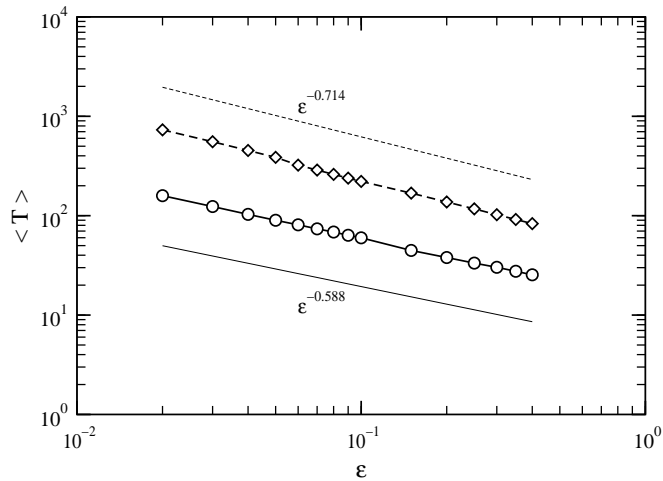}}}
}
\caption{
Variation of the mean first passage time with
$\varepsilon$ for a free particle on an interval of length
$L=3$ and two different power laws. Circles: $\beta = 0.3$;
Diamonds: $\beta = 0.6$. The exponents
obtained from fitting the simulation results are $0.61$ and $0.72$. The
auxiliary lines correspond to the theoretical exponents
($1/(2-\beta)$), which are $0.588$ and $0.714$, respectively.
}
\label{fig:tempe}
\end{figure}

\begin{figure}[htb]
\vspace{0.5cm}
\begin{center}
\mbox{\scalebox{.35}{\includegraphics[angle=-90]{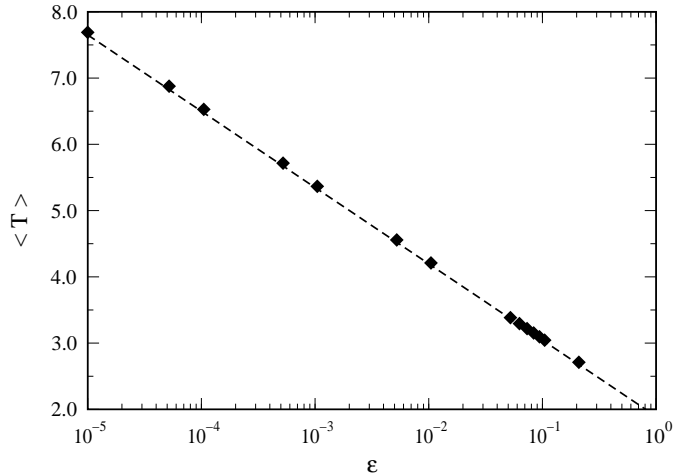}}}
\end{center}
\caption{Mean first passage time out of an unstable state
as a function
of $\varepsilon$ for fixed $\beta = 0.5$.  The
dashed line is the
theoretical prediction from Eq.~(\protect \ref{MFPT}\protect ) 
and the statistical average is over 5000 realizations. 
}
\label{fig:mfpt.uns1}
\end{figure}
\begin{figure}[htb]
\vspace{0.5cm}
\begin{center}
\mbox{\scalebox{.35}{\includegraphics[angle=-90]{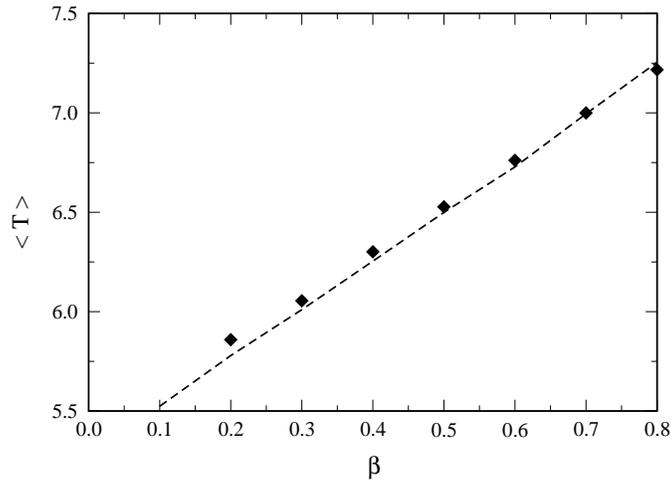}}}
\end{center}
\caption{ Mean first passage time out of an unstable state as function
of the power law exponent $\beta$ for fixed $\epsilon = 0.0001$. The
dashed line is the
theoretical prediction from Eq.~(\protect \ref{MFPT}\protect ) 
and the statistical average is over 5000 realizations. 
}
\label{fig:mfpt.uns2}
\end{figure}

\section{Superdiffusion of a free particle}
\label{superdiffusion}

The first problem we study is of interest in the analysis of the
mean exit time of a free Brownian particle from a domain bounded by
absorbing walls.  We consider a free particle that
moves in one dimension under the influence of a long range correlated noise
until it covers, on average, the distance to one of the absorbing 
barriers. 

A dynamical equation that describes this type of dynamics is the
very simple stochastic differential equation (Langevin equation)
\begin{equation}
\frac{d x}{dt} = \eta (t),
\label{free}
\end{equation}
where $\eta(t)$ is a Gaussian 
power law-correlated random noise as described in section~\ref{definenoise}.
Note that here and in all subsequent examples we ``start the clock"
at $t=0$, that is, in addition to specifying an initial value or
distribution of $x$ at $t=0$ we also assume that the noise is ``turned on"
at that time.  Otherwise one would need to be concerned about the role of
the time $t=0$ in the evolution of the noise, which in turn would affect
the description of its correlation function. 

We are particularly interested in the dependence of the mean
arrival (absorption) time on the parameters $\beta$ and $\epsilon$ and on
the length $2L$ of the interval.
The essential scaling features
(exponents) of the dependences on these parameters
can be obtained from a formal integration of Eq.(\ref{free}), 
which immediately leads to the mean squared displacement relation
\begin{equation}
\langle x^2(t) \rangle \sim \varepsilon \,t^{2 -\beta},
\label{formal}
\end{equation}
characteristic of {\em superdiffusive} behavior~\cite{Romero99}.
Superdiffusion here arises because the arrival at the boundaries is more
likely to occur via essentially ballistic motion than in the white noise
case, where frequent reversals of $dx/dt$ cause the net displacement to be
slow.
The time at which the mean squared displacement is proportional to
$L^2$ can be obtained by inverting this expression.  The difference between
this time and the mean time $\langle T \rangle$ for {\em first} arrival at an
absorbing boundary is due to recrossings of the boundary and only
affects the prefactors, not the exponents. That is, we can use
Eq.~(\ref{formal}) to write
\begin{equation}
\langle T \rangle  \sim \varepsilon^{-1/(2-\beta)}\, L^{2/(2-\beta)}.
\label{freeT}
\end{equation}
Note that the (Gaussian) statistics of the noise play no role in this
result, which is determined entirely by the correlation function of the
noise.

In order to test this theoretical prediction we have performed numerical 
simulations for different values of the interval length $2L$, the
exponent $\beta$, and the coefficient $\varepsilon$. 
Throughout this paper all Langevin equations are numerically
integrated using the Heun method, which is an
extension of a second order Runge-Kutta  algorithm for stochastic
differential equations~\cite{Toral95}.  The results presented below are
averages over $2000$ realizations.

A comparison of the theoretical exponent ($2/(2-\beta)$) in
Eq.~(\ref{freeT})
and the simulation results is summarized in Table~I.
The agreement is clearly excellent.
In Fig.~\ref{fig:xt} the average exit time is
presented as function of the interval length on a log-log plot.
The exit time $\langle T \rangle$ follows the predicted power law
behavior with the slopes of Table~I.
In the figure we also show
the result for white noise with its characteristic exponent of $2$.

Fig. \ref{fig:tempe} shows the dependence of the mean first
passage time on the noise intensity parameter $\varepsilon$. The
simulation results are in excellent agreement with the 
predictions of Eq.~(\ref{freeT}).

\section{Decay of an unstable state}
\label{decay}

The decay of unstable states is of course easily triggered by any
disturbance, and in many physical instances the decay is caused by
fluctuations.  
In its simplest rendition this process can be modeled by a particle
initially placed at the top ($x=0$) of an inverted parabolic potential and
subject to fluctuations,
\begin{equation}
\frac{d x}{dt} =  x -x^3 + \eta (t), \qquad x(0) = 0.
\label{unstable1}
\end{equation}
The decay at the early stages is dominated by the linear term and the
fluctuations, so that in this regime the Langevin equation can be further
simplified to~\cite{sancho89}
\begin{equation}
\frac{d x}{dt} =  x + \eta (t). 
\label{unstable}
\end{equation}
Formal integration of this linear equation immediately yields
\begin{equation}
x(t) = h(t) e^t, \qquad \qquad h(t) \equiv \int_0^{t} dt' e^{-t'} 
\eta(t').
\label{int}
\end{equation}
The mean first passage time to, say, $x=\pm 1$ is identified as the time at
which $\langle x^2(t)\rangle =1$. Recrossings are even less relevant here than
in the previous section; larger values of $x$ require consideration of
the nonlinear term in the evolution equation.  If the time of interest is
$t\gg 1$ (a condition that will be seen to be satisfied by the mean first
passage time) then the upper limit in the integral in Eq.~(\ref{int}) can
be extended to infinity, that is, we set
\begin{equation}
x^2(t) = h^2 e^2t, \qquad \qquad h = \int_0^{\infty} dt' e^{-t'}
\eta(t').
\label{int2}
\end{equation}
Thus the fluctuations $h^2$ simply play the role of a random initial
condition on $x^2(t)$.  The random variable $h$ has zero mean and
variance $\sigma^2$ given by \cite{caceres},
\begin{eqnarray}
\sigma^2  = \langle h^2\rangle =  \langle \int_0^{\infty} dt'
e^{-t'}\,\eta(t') 
\int_0^{\infty} dt'' e^{-t''}\,\eta(t'') \rangle = 
\nonumber
\\
\int_0^{\infty} e^{-2t'}dt' \int_0^{\infty} dt'' e^{-t''} \gamma(s) =
\frac{\epsilon}{2}\,t_0^{\beta} e^{t_0} \Gamma ( 1-\beta, t_0).
\label{eq:sigma}
\end{eqnarray}
The stationarity of the noise has been used to calculate the
variance; $\Gamma ( 1-\beta, t_0)$ is the incomplete Gamma function.  

The mean first passage time to $x^2= 1$ for this process  
for a generic colored noise was calculated in Ref.~\cite{sancho89}:
\begin{equation}
\langle T\rangle  = - \frac{1}{2} \ln \left( \sigma^2 \right)
+\frac{\gamma}{2},
\label{MFPT}
\end{equation}
where $\gamma = 0.57721...$ is the Euler constant. This result together
with (\ref{eq:sigma}) gives an explicit expression for the
mean first passage time.  Figures~\ref{fig:mfpt.uns1} and
\ref{fig:mfpt.uns2} show excellent agreement
between the theory and numerical simulations for the dependence of $\langle
T \rangle$ on $\varepsilon$ and on $\beta$.

\section{The Barrier Crossing Problem}
\label{barrier}

A problem that is more complex than the essentially linear ones posed
so far is the ``barrier crossing problem."  Here we consider the diffusion
of an overdamped  particle in a double-well potential.  The mean first
passage time of interest is the time it takes the particle to go from
one of the potential minima to the other when the transition is driven
by power-law-correlated Gaussian fluctuations.  The particle
dynamics is modeled by the following Langevin equation:

\begin{figure}[htb]
\centerline{\psfig{file=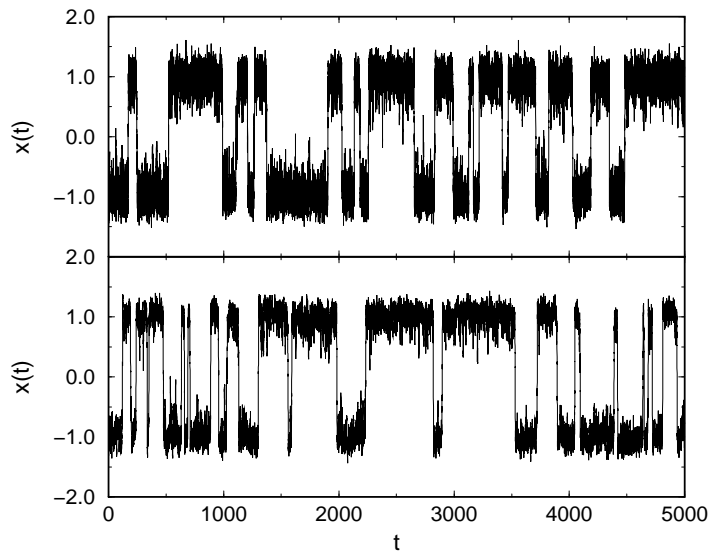,width=9cm}}
\caption{
Typical trajectories of a particle in the double well 
potential (\protect\ref{well}).
Top panel: Gaussian white noise with $D = 0.0725$,
$\langle T \rangle = 162.38$. 
Bottom: Gaussian power-law noise with $\beta = 0.5$, $\epsilon = 1.15$,
$\langle T \rangle = 162.87$. 
}
\label{fig:xtime}
\vspace*{0.5in}
\end{figure}
\begin{figure}[htb]
\centerline{
\mbox{\scalebox{.35}{\includegraphics[angle=-90]{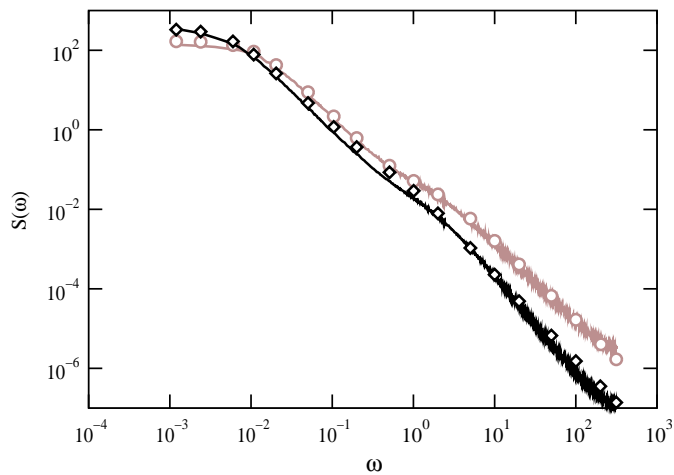}}}
}
\caption{
Power spectra of the trajectories shown in
Fig.~\protect \ref{fig:xtime} \protect.
Circles: white noise with $D = 0.0725$ and
$\langle T \rangle = 162.38$. 
Diamonds: power-law noise with $\beta = 0.5$, $\epsilon = 1.15$ and
$\langle T \rangle = 162.87$. 
The theoretical predictions from Eq.~(\protect \ref{spectrum})\protect
are also included.
}
\label{fig:pwxt1}
\end{figure}
\begin{figure}[htb]
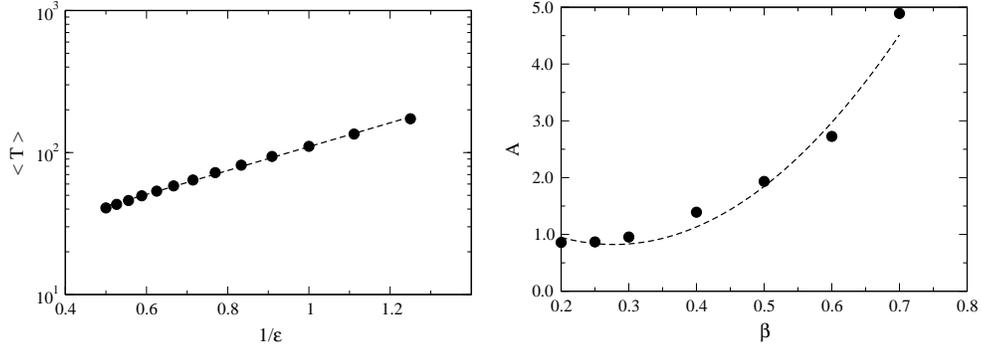

\centerline{
\mbox{\scalebox{.25}{\includegraphics[angle=-90]{TvsEpsilon.eps}}}
\quad
\mbox{\scalebox{.25}{\includegraphics[angle=-90]{AvsBeta.eps}}}
}
\caption{
First panel:Typical dependence on $\varepsilon$ of the mean
first passage time from
one well to the other. The other parameters are $\beta = 0.5$,
$\omega_0 = 0.0001$, and $\Delta t = 0.02$.  The dashed line
corresponds to $A = 1.933$.  Second panel: 
Activation parameter $A$
as a function of $\beta$ (the value $A=1.933$ at $\beta=0.5$ is the result
of the slope in the first panel.
The dashed curve is the best numerical fit with a quadratic form,
$A(\beta) = 2.42 + 11.5 (1.8 \beta^2 - \beta )$.
}
\label{quasikramers1}
\vspace*{0.2in}
\end{figure}
\begin{figure}[htb]
\vspace{.5cm}
\begin{center}
\mbox{\scalebox{.35}{\includegraphics[angle=-90]{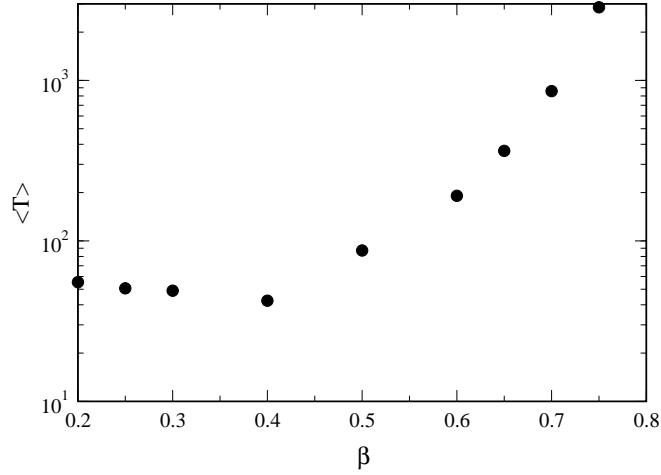}}}
\end{center}
\caption{Mean transition time from one well to the other
as a function of $\beta$.  The parameters used here are
$\varepsilon = 1.15$, $\omega_0 = 0.0001$, $\Delta t = 0.02$. 
}
\label{versusbeta}
\end{figure}
\begin{figure}[htb]
\begin{center}
\mbox{\scalebox{.35}{\includegraphics[angle=-90]{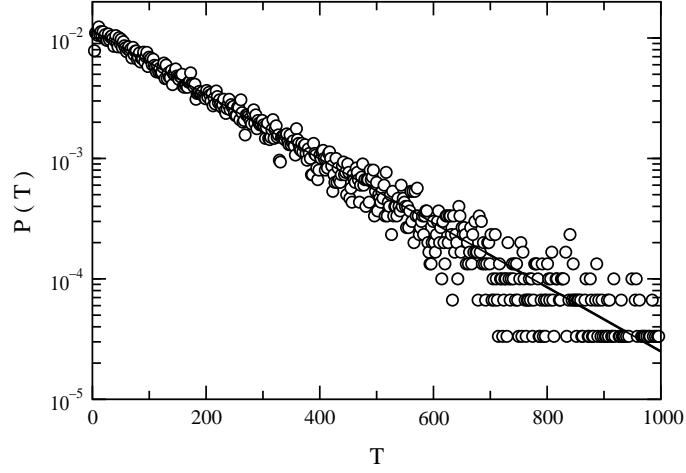}}}
\end{center}
\caption{Transition time probability distributions for 
white noise with $D = 0.0725$. 
The circles are the results of the simulation
and the continuous line is the functions
Eq.~(\protect\ref{exponential}) with
$\langle T \rangle$ obtained from the simulation data.}
\label{P(T)1}
\vspace*{0.4in}
\end{figure}
\begin{figure}[htb]
\centerline{
\mbox{\scalebox{.35}{\includegraphics[angle=-90]{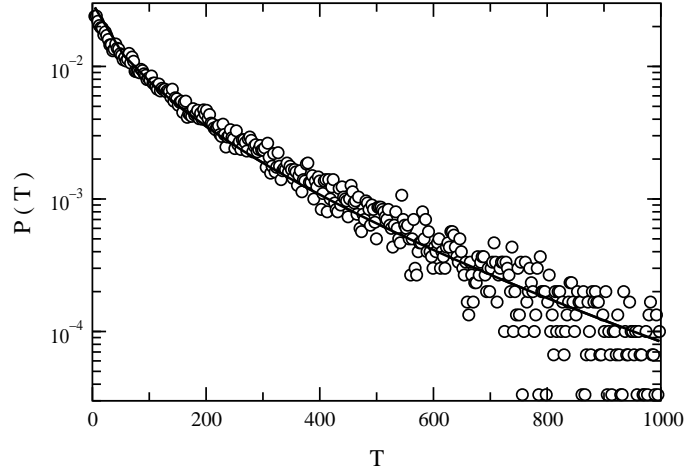}}}
}
\caption{Transition time probability distribution for the
power law case with
$\beta = 0.5, \epsilon=1.15$. 
The circles are the results of the simulation
and the continuous line is the function
Eq.~(\protect\ref{stretched}) with
$T_s$, and $\theta$ obtained from the simulation data.}
\label{P(T)2}
\end{figure}
\begin{figure}[htb]
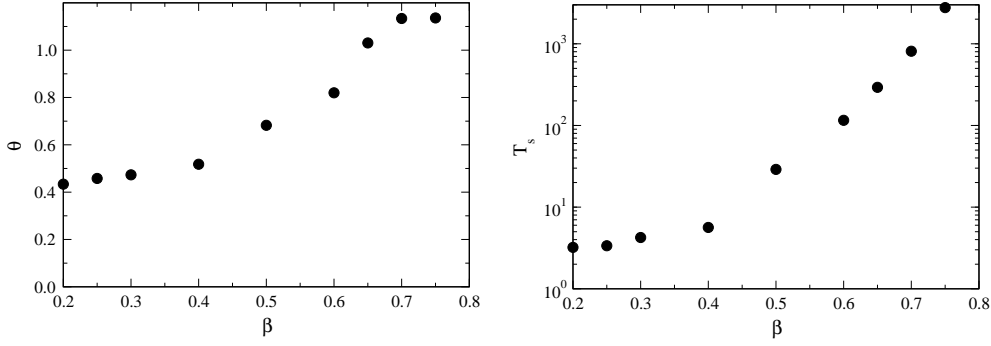

\vspace{0.5cm}
\centerline{
\mbox{\scalebox{.25}{\includegraphics[angle=-90]{TethavsBeta.eps}}}
\quad
\mbox{\scalebox{.25}{\includegraphics[angle=-90]{T_svsBeta.eps}}}
}
\caption{First panel: Characteristic parameter $\theta$ of the stretched
exponential distribution as a function of $\beta$.  Secon panel:
characteristic parameter $T_s$.  The values of the
other parameters are $\varepsilon=1.15$ and $\Delta t = 0.02$.
}
\label{tethavsbeta1}
\vspace*{0.2in}
\end{figure}
\begin{figure}[htb]
\centerline{
\mbox{\scalebox{.35}{\includegraphics[angle=-90]{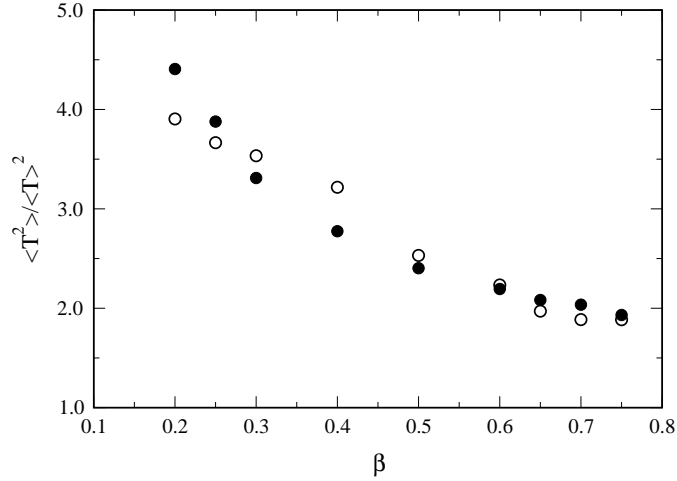}}}
}
\caption{
Dispersion of the first passage time distribution as a function of
$\beta$ (full circles) compared with the proposed analytic form
Eq.~(\ref{T1T2}) (empty circles).
}
\label{T2TvsBeta}
\end{figure}

\begin{equation}
\frac{dx}{dt} = - \frac{d V(x)}{dt} + \eta (t),
\label{well}
\end{equation}
where the double-well potential is,  
\begin{equation}
V(x) = - \frac{1}{2} x^2 + \frac{1}{4} x^4 .
\end{equation}
Our simulations follow the dynamics of each particle
starting in the left well at $x(t=0)=-1.0$ until it arrives at the
right well at $x(t=T)=1.0$ for the first time.  We then calculate the
mean first passage time $\langle T \rangle$, that is, the average
of $T$ over many ($10,000$) realizations.

Two questions are of interest: (1) If two noises $\eta(t)$, one
$\delta$-correlated and the other power-law correlated, lead to the same
average transition time from one well to the another, are there other
properties that allow a clear distinction between them?  (2) For a
power-law correlated noise $\eta(t)$, how does the transition time depend
on the correlation function parameters?  We address both of these questions
below.

The numerical barrier crossing problem is fundamentally different
from the problems considered in the previous section. In those, no
matter how small or persistent a realization of the noise $\eta(t)$,
the process $x(t)$ will eventually reach one of the boundaries of
interest. How long a simulation must run in order to make sure that
all processes have reached the boundaries is essentially a matter of 
insuring that those with the lowest value of $\eta$ as given in the
discretization scheme arrive there.  In the barrier crossing problem, on
the other hand, small-$\eta$ realizations will remain in one
well until the value of $\eta$ changes to a sufficiently high value
(easy to estimate~\cite{reviews}) so as to effectively eliminate the barrier.  
The time it takes to effect this change is of course large (infinite on
average) for the fluctuations considered herein.   In our simulations we
have carefully chosen run times that insure that all $10,000$ particles in
our ensemble have made the passage from one well to the other.  As we
decrease $\varepsilon$ (thus weakening the noise) and/or increase $\beta$
(thus in effect also weakening the noise), the simulation time has to be
increased accordingly.  Note that these remarks address the {\em simulation}
problem.  A complete theory of this process would have to deal with 
the relative magnitudes of a number of infinities (noise intensity, noise
correlation, and passage over the barrier of the slowest processes).

Fig. \ref{fig:xtime} shows two typical trajectories of the 
barrier crossing process for two different Gaussian noises. One
is white noise, the other is power-law correlated, and the parameters
have been chosen so that the mean first
passage times for both are essentially the same.  The effect of the long
range correlations is certainly not evident from these trajectories.
This observation suggests a more detailed study of the dynamical
properties.  The first passage time distribution and, in
particular, the mean first passage time for the Gaussian white
noise case, are well known analytically (see~\cite{reviews} and
the many original references therein).  Approximate theories for
exponentially correlated noise also abound~\cite{reviews}, but to
our knowledge there are no results for the infinite correlation time case.
To deal with this problem we note that the trajectories
(in either the white or the correlated noise cases) consist of two distinct
components characterized by two different time scales.  One component
describes the
rapid fluctuations in each well, and the other describes the much
slower switching events between the two wells.  We implement this
observation by writing the 
trajectory $x(t)$ as a sum of these two contributions: 
\begin{equation}
x(t) = x_D(t) + x_B(t),
\end{equation}
where $x_B(t)$ represents the rapid random motion inside each
well, and 
$X_D(t)$ is a dichotomous random process between the values $x =
1$ and $x=-1$,
with a characteristic time controlled by the mean transition 
time $\langle T \rangle$.
The dynamics of $x_B(t)$ can be approximated by expanding
Eq.~(\ref{well}) around the minimum of one of the wells, e.g.
as $x(t) = 1 + x_B(t)$ in the well. 
Then $x_B(t)$ evolves according to the Langevin equation
\begin{equation}
\frac{d x_B}{dt} = - 2 x_B + \eta(t),
\label{xB}
\end{equation}
which evolves on a characteristic time scale $\tau = 0.5$ that is much 
smaller than $\langle T \rangle$.

With this decomposition, the  
power spectrum of $x(t)$, denoted by $S(\omega)$, is just the sum 
of the spectra associated with $x_D(t)$ and $x_B(t)$.  These can be 
calculated explicitly:
\begin{equation}
S(\omega) = \frac{4\langle T \rangle }{ 4 + 
\omega^2 \langle T \rangle^2}
+ \frac {\tau^2\gamma(\omega)}{ 1 +
\omega^2\tau^2}.
\label{spectrum}
\end{equation} 
The first contribution is
the spectrum of a dichotomous process governed by the time scale 
$\langle T \rangle$,
and the second contribution comes from the exact solution of Eq. (\ref{xB})
with the noise spectrum $\gamma(\omega)$. 
In the white noise case $\gamma(\omega)= 2D$, 
and for the long ranged noise the spectrum is obtained
from Eq.~(\ref{eta}) in the continuum limit,
\begin{equation}
\gamma(\omega) = 
\frac{\epsilon \beta \pi \omega^{-\beta}_{max}}{( \omega +
\omega_0)^{1 - \beta}}
\end{equation}
It is interesting to note that the main difference between the white
and power law spectra appear in the high-frequency (short-time) regime. 
In the white noise case, $S(\omega) \sim \omega^{-2}$, whereas in the
correlated case we observe the more rapid decay
$S(\omega) \sim \omega^{-(3-\beta)}$. 

Figure~\ref{fig:pwxt1} shows the spectra for the two trajectories
shown in Fig.~\ref{fig:xtime} as well as the prediction (\ref{spectrum})
for each.  
As noted earlier, the parameters have been chosen so as to lead to the same
mean first passage time $\langle T \rangle$, but other than that
there are no adjustable parameters.  The agreement between our predictions
and the numerical results is quite good in both cases. The exponent
obtained from the numerical data in the high frequency regime for
the correlated noise case is 2.49, which agrees with the theoretical
prediction of 2.5.  It is also clear that a distinction between the two
types of fluctuations is difficult even at this level of detail for
a given mean transition time.

In the above analysis $\langle T \rangle$ is the adjustable parameter, that
is, we have fixed $\beta$ and $\varepsilon$ so as to yield a particular
value of $\langle T \rangle$.  The dependence of this quantity on the
model parameters $\varepsilon$ and $\beta$ is yet to be explored.

The first panel in Fig.~\ref{quasikramers1} is a plot of $\log \langle T
\rangle$ as a function of $1/\varepsilon$.  The straight line behavior
shows that the transition time is described by a law of the Kramers form,
\begin{equation}
\langle T \rangle \sim \exp\left( \frac{A}{\varepsilon}\right).
\label{kramers}
\end{equation}
Fitting the simulation data in the figure with a straight line
we obtain $A=1.933$, very
far from the value $A= 2/27$ obtained for exponentially correlated
noise~\cite{reviews}.
We stress that the ``activation parameter" $A$ here depends on $\beta$,
that is, a plot exactly like that of the first panel
Fig.~\ref{quasikramers1} but for a
different value of $\beta$ leads to a different value of
$A$.  The values of $A$ vs $\beta$ collected in this way are shown in 
the second panel in Fig.~\ref{quasikramers1}.
In contrast, for noises with a finite (albeit large)
correlation time $A$ is essentially a constant independent of the
noise intensity and of the correlation time.  Comparison with those results
is therefore not appropriate.  

The dependence of the mean transition time on $\beta$ qualitatively
parallels that of the activation parameter but is even more complex because
it is non-monotonic.  The typical behavior is shown in
Fig.~\ref{versusbeta}.  Although we have no
detailed theory to account for this non-monotonicity, a qualitative
explanation is possible in terms of two competing effects, one that 
causes a decrease in $\langle T \rangle$ with increasing $\beta$ and the other
that causes $\langle T \rangle$  to increase with increasing $\beta$.  
At very small $\beta$
the mean first passage time is dominated by those realizations of very slow
passage for which $\eta(0)$ is not large enough to cause a transition and a
very long wait is involved before $\eta(t)$ changes to a value that is
sufficiently large.  With increasing $\beta$ the wait for change decreases,
thus leading to faster passage.  On the other hand, increasing $\beta$ in
effect {\em decreases} the intensity of the noise.   To see this suppose
that we calculate an effective intensity by integrating the correlation
function only up to a maximum time $t_{max}$ ($\gg t_0$), the running
time of our simulations:
\begin{eqnarray}
\int_0^{t_{max}} dt \gamma(t) &=& \frac{\varepsilon t_0}{1-\beta}
\left[\left(1+\frac{\displaystyle t_{max}}{\displaystyle t_0}\right)^{1-\beta}
-1\right] \nonumber\\ [8pt]
&=& \frac{\varepsilon t_0}{1-\beta}
\left(\frac{\displaystyle t_{max}}{\displaystyle t_0}\right)^{1-\beta}
\left[1+{\cal O}\left(\frac{\displaystyle (1-\beta)t_0}
{\displaystyle t_{max}}\right)\right].
\label{integr}
\end{eqnarray}
The $1-\beta$ in the exponent of the large ratio $t_{max}/t_0$ 
dominates the $1-\beta$-dependence in the denominator, so that
for a fixed $t_{max}$ this decreases rapidly with increasing
$\beta$.
This lower effective intensity of the noise leads to slower passage. 
We conjecture that the interplay of the two effects leads to the
non-monotonic behavior observed in Fig.~\ref{versusbeta}, although the
specific value and specific location of the minimum depend on
$\varepsilon$.

In order to verify this minimum and to exhibit  
more information on the statistics of $\langle T \rangle$, 
we have also obtained numerical results for  the probability
distribution $P(T)$ of the exit times (of which $\langle T \rangle$ is the
average).  In Figs.~\ref{P(T)1} and \ref{P(T)2} we have plotted
the numerical histogram
of $P(T)$ for white noise and for a power-law correlated noise, again
chosen so as to correspond to the same means.  The scatter of numerical
points around the analytic forms discussed below (solid lines) is quite
large and could be decreased by doing many more realizations for much longer
times.

In the case of white noise the first passage time distribution is known to
be exponential~\cite{risken,FPfpt},
\begin{equation}
P(T) = \frac{1}{\langle T \rangle} \exp\left({-\frac{T}{<T>}}\right).
\label{exponential}
\end{equation} 
Our numerical results agree with this result.  For the power law case
we conjecture a stretched exponential distribution for the transition time,
\begin{equation}
P(T) = \frac{\theta}{T_s \Gamma(1/\theta)}
\exp \left( -\left[ \frac{T}{T_s} \right] ^{\theta} \right),
\label{stretched}
\end{equation}
where the prefactor has been chosen to ensure normalization
($\Gamma(z)$ is the Gamma function) and
where the parameters $\theta$ and $T_s$ are obtained from the
simulation data.  The mean first passage time in terms of these parameters
is 
\begin{equation}
\langle T \rangle = T_s\frac {\Gamma(2/\theta)} {\Gamma(1/\theta)} =
\frac{T_s}{\sqrt{2\pi}}
2^{\frac{2}{\theta}-\frac{1}{2}}\Gamma(\frac{1}{\theta}+\frac{1}{2}).
\label{independent}
\end{equation}
We do not display the $\varepsilon$-dependence of the parameters because
it is the $\beta$ dependence of the
first passage time that requires further understanding.
Our fitting procedure for the case of $\epsilon = 1.15$ and $\Delta t =
0.02$ gives the $\beta$-dependences observed in Fig.~\ref{tethavsbeta1}.
Although the dependences of the parameters on $\beta$ are monotonic, the
dependence of $\langle T \rangle$ on $\theta$ is not monotonic 
(note that $T_s$ also varies with $\theta$)
and gives rise to the observed minimum.
Using the data shown in Fig.~\ref{tethavsbeta1}
to construct $\langle T
\rangle$ according to Eq.~(\ref{independent}) one again obtains the minimum
as a function of $\beta$ observed in Fig.~\ref{versusbeta}.
Another point worth noting is the approach of the exponent $\theta$ to
unity with increasing $\beta$ that is seen in Fig.~\ref{tethavsbeta1}.
The exponent is exactly unity for Gaussian white noise and would also
be unity for $\beta>2$.

From the proposed distribution (\ref{stretched}) one can evaluate the
second moment of the first passage time distribution.  We obtain
\begin{equation}
\frac{ \langle T^2 \rangle}{\langle T \rangle^2} = \frac{\Gamma(1/\theta) \,
\Gamma(3/\theta)}{\Gamma(2/\theta)^2}\,, 
\label{T1T2}
\end{equation}
For white Gaussian noise this ratio is equal to $2$.  We see in 
Fig.~\ref{T2TvsBeta} that for correlated noise the ratio is greater than
$2$ reflecting the greater width of the stretched exponential distribution,
but with increasing $\beta$ the approach toward $2$ is evident. The figure
shows the moment ratio $\langle T^2\rangle/\langle T\rangle^2$
obtained from numerical simulations (full circles) and from the analytic
expression Eq.~(\ref{T1T2}).

\section{Summary and Conclusions}
\label{conclusions}

We have presented a numerical study and derived (in some cases) or
conjectured (in others) analytic results for the first passage
time problem in systems driven by long-range correlated noises.  We
specifically considered {\em Gaussian} noise, as opposed to noises with
L\'{e}vy or other long-tailed distributions, or noises defined in terms of
long waiting times between events, that may also lead to long-range
correlations.  There is an extensive literature on these non-Gaussian
problems~\cite{klafter}, but very little on the Gaussian
counterpart~\cite{makse96,nises,Romero99}.  We have considered highly
correlated noise with an inverse power law form for the correlation
function, Eq.~(\ref{powercorr}),
with $0<\beta<1$.  Not only does this noise have no finite correlation
time, but the correlation function is not integrable.  The parameters of
the problem are the exponent $\beta$ and the noise control parameter
$\varepsilon$.

First we discussed a numerical algorithm to generate this type of noise.
This is in itself not a trivial problem.  Then we considered several
systems driven by Gaussian power-law correlated noises with a special
focus on the problem of first passage to a prescribed boundary or set of
boundaries.  Two of these problems, the arrival of a free particle in one
dimension at either end of a finite interval, and the decay of an unstable
state, admit straightforward analytic solution.  The dependence of the
respective mean first passage times on the noise parameters were
calculated explicitly and compared favorably with numerical simulations.  

Our final application is more complex: Here we considered the first
passage of a particle that evolves in a double
well potential from one well, over the barrier, to the other well.  
We specifically considered two issues.  One is a comparison of the
trajectories of such a particle subject to Gaussian white noise and to
Gaussian power-law correlated noise, with parameters chosen in such a way
that the mean first passage times for crossing from one well to the other
in both cases is the same.  It is difficult to find a clear difference that
would reveal in an experiment which type of noise the particle was
subjected to.  Only the high frequency decay of the respective power
spectra shows some differences.  The other issue we explored is the
dependence of the first passage time distribution on the parameters $\beta$
and $\varepsilon$.  We found that the dependence of the mean first passage
time on the noise control parameter is of the usual ``activated" form, but
with an effective activation parameter that depends on $\beta$.  We also
found that the mean first passage time exhibits a minimum as a function of
$\beta$, which we explain on the basis of competing mechanisms, one of which
leads to an increase of the mean first passage time with increasing $\beta$
and the other to a decrease.  This non-monotonicity is a signature
characteristic of these fluctuations.  We also found that a stretched
exponential form for the first passage time distribution describes the
numerical results, with a stretched exponent $\theta$
and a characteristic time $T_s$ that both depend on $\beta$.   This
stretched exponential form is consistent with the non-monotonic behavior of
the mean first passage time, and also gives results in agreement with the
simulation outcomes for the second moment of the first passage time
distribution.  
The stretched exponential first passage time distribution, and the
dependence of the parameters of this distribution on the noise parameters
$\varepsilon$ and $\beta$, are features that have no analog in Gaussian
noises with a finite correlation time.

\section*{Acknowledgements}

The work was supported in part by the U.S. Department of Energy under Grant 
No. DE-FG03-86ER13606, by FONDECYT, Chile under Grant 1010988,
by the Millenium Project under Grant P99-135F, and by the
Comisi\'on Interministerial de 
Ciencia y Tecnolog\'{\i}a (Spain) Project No. BFM2000-0624.


\begin{thebibliography}{99}
\bibitem{note} In Ref.~\protect\cite{makse96}\protect  a correlation
function with the same asymptotic power law behavior but with an analytic
Fourier transform is used.  Since the statistical properties of interest
are determined primarily by the asymptotic behavior of the correlation
function, this choice would lead to similar results in our applications.
However, we are interested in exploring the numerical demands of a less
restricted assumption.
\end{thebibliography}
\end{document}